\documentclass[aip,graphicx,reprint]{revtex4-1}
\usepackage{graphicx}
\usepackage{hyperref}
\hypersetup{colorlinks=true,citecolor=blue,linkcolor=blue,urlcolor=blue}
\usepackage{amsmath}
\usepackage{times}
\usepackage{subfigure}
\begin{document}


\title{Integrating MBE materials with graphene to induce novel spin-based phenomena} 



\author{Adrian G. Swartz}
\affiliation{Department of Physics and Astronomy, University of California, Riverside,
CA 92521, USA}

\author{Kathleen M. McCreary}
\affiliation{Department of Physics and Astronomy, University of California, Riverside,
CA 92521, USA}

\author{Wei Han}
\affiliation{Department of Physics and Astronomy, University of California, Riverside,
CA 92521, USA}

\author{Jared J. I. Wong}
\affiliation{Department of Physics and Astronomy, University of California, Riverside,
CA 92521, USA}

\author{Patrick M. Odenthal}
\affiliation{Department of Physics and Astronomy, University of California, Riverside,
CA 92521, USA}

\author{Hua Wen}
\affiliation{Department of Physics and Astronomy, University of California, Riverside,
CA 92521, USA}

\author{Jen-Ru Chen}
\affiliation{Department of Physics and Astronomy, University of California, Riverside,
CA 92521, USA}

\author{Yufeng Hao}
\author{Rodney S. Ruoff}
\affiliation{Department of Mechanical Engineering and the Texas Materials Institute, The University of Texas at Austin, Austin TX 78712}

\author{Jaroslav Fabian}
\affiliation{Institute for Theoretical Physics, University of Regensburg, D-93040 Regensburg, Germany}

\author{Roland K. Kawakami}
\affiliation{Department of Physics and Astronomy, University of California, Riverside,
CA 92521, USA}


\date{\today}

\begin{abstract}
Magnetism in graphene is an emerging field that has received much theoretical attention. In particular, there have been exciting predictions for induced magnetism through proximity to a ferromagnetic insulator as well as through localized dopants and defects.  Here, we discuss our experimental work using molecular beam epitaxy (MBE) to modify the surface of graphene and induce novel spin-dependent phenomena. First, we investigate the epitaxial growth the ferromagnetic insulator EuO on graphene and discuss possible scenarios for realizing exchange splitting and exchange fields by ferromagnetic insulators. Second, we investigate the properties of magnetic moments in graphene originating from localized $p_z$-orbital defects (i.e.~adsorbed hydrogen atoms). The behavior of these magnetic moments is studied using non-local spin transport to directly probe the spin-degree of freedom of the defect-induced states. We also report the presence of enhanced electron $g$-factors caused by the exchange fields present in the system. Importantly, the exchange field is found to be highly gate dependent, with decreasing $g$-factors with increasing carrier densities.
\end{abstract}


\maketitle 

\section{Introduction}
Magnetism in graphene is an emerging field that has received much theoretical attention \cite{Yazyev:2010,Palacios:2008,Yazyev:2007,Haugen:2008,Semenov:2007,Yang:2013}. In particular, there have been exciting predictions for induced magnetism through proximity to a ferromagnetic insulator \cite{Semenov:2007,Haugen:2008} as well as through localized dopants and defects \cite{Yazyev:2010}. While these systems appear to be very different, induced magnetic phenomena in both systems relies on the principle of the exchange interaction. In the case of ferromagnetic insulators in contact with graphene, atomic orbital overlap at the interface is expected to induce a spin-splitting in the graphene layer \cite{Semenov:2007,Haugen:2008,Yang:2013}. Alternatively, on a local scale, magnetism can be induced in graphene through dopants and defects \cite{Yazyev:2010,Nair:2012,Mccreary:2012}. In this scenario, adsorbates or lattice vacancies effectively remove a $p_z$-orbital from the graphene band structure. This is known to create a localized defect state near the Fermi energy \cite{Pereira:2006,Huang:2009}. Due to the Coulomb interaction, the defect state is spin split leading to a spin-1/2 populated quasi-localized defect state \cite{Yazyev:2007,Palacios:2008,Yazyev:2010}. As both types of induced magnetism rely on the exchange interaction which requires atomic scale overlap of wavefunctions, materials control on this scale is absolutely necessary for realizing these new behaviors. This is possible through molecular beam epitaxy (MBE) which typically has very slow deposition rates ($\sim\text{\AA}$/min) allowing for high purity growth control in the monolayer or sub-monolayer range.

Here, we discuss our work on induced spin-dependent phenomenon in graphene. First, in section \ref{sec:EuO} we discuss the epitaxial growth of the ferromagnetic insulator EuO on graphene and discuss possible scenarios for realizing exchange splitting and exchange fields by ferromagnetic insulators. Second, in section \ref{sec:defects} we discuss our ongoing efforts on induced magnetism in graphene through localized $p_z$-orbital defects. Using non-local spin transport to probe the spin-degree of freedom of the quasi-localized states, we have observed the presence of these magnetic states. These states are coupled via exchange to the injected spin current which experiences an effective exchange field due to the presence of the magnetic moments. In particular, through Hanle spin precession measurements, we observe enhanced electron $g$-factors caused by the exchange fields present in the system. Importantly, the exchange field observed is highly gate dependent, with decreasing $g$-factors with increasing carrier densities. Lastly, we discuss an important issue for the field of spintronics related to the analysis of Hanle spin precession data in the case of an unknown effective electron $g$-factor. 

\section{Realization of Epitaxial Ferromagnetic Insulators on Graphene By reactive MBE}
\label{sec:EuO}
EuO is discussed in the literature as a possible ferromagnetic insulator for inducing exchange by proximity \cite{Haugen:2008,Asano:2008,Michetti:2010,Michetti:2011,Zhang:2011,Yang:2013}. This is understandable since EuO has excellent intrinsic properties including large magnetization per unit volume, is highly resistive in the insulating state, and fits the ideal model for an isotropic Heisenberg ferromagnet \cite{Mauger:1986}. This spin splitting was originally predicted to be approximately 5 meV based on measurements by Tedrow dating back to 1986 \cite{Haugen:2008,Tedrow:1986}. Very recently, this has been revisited by density functional calculations that has indicated that the induced exchange splitting can be as large as 35 meV \cite{Yang:2013}. Further, there have been many fantastic theoretical suggestions for possible novel device applications including controllable magnetoresistance \cite{Haugen:2008,Semenov:2008a,Zou:2009,Yu:2011}, gate tunable manipulation of spin transport \cite{Semenov:2007,Michetti:2010}, gate tunable exchange bias \cite{Semenov:2008b}, spin transfer torque \cite{Zhou:2010,Yokoyama:2011}. Also, such exchange splitting is a key requirement for realizing the quantized anomalous Hall effect in graphene systems \cite{Qiao:2010,Tse:2011}. To date, exchange splitting induced by a ferromagnetic insulator in contact with graphene has not been demonstrated. 

\begin{figure}[t]
\centering
\includegraphics[scale=1.05]{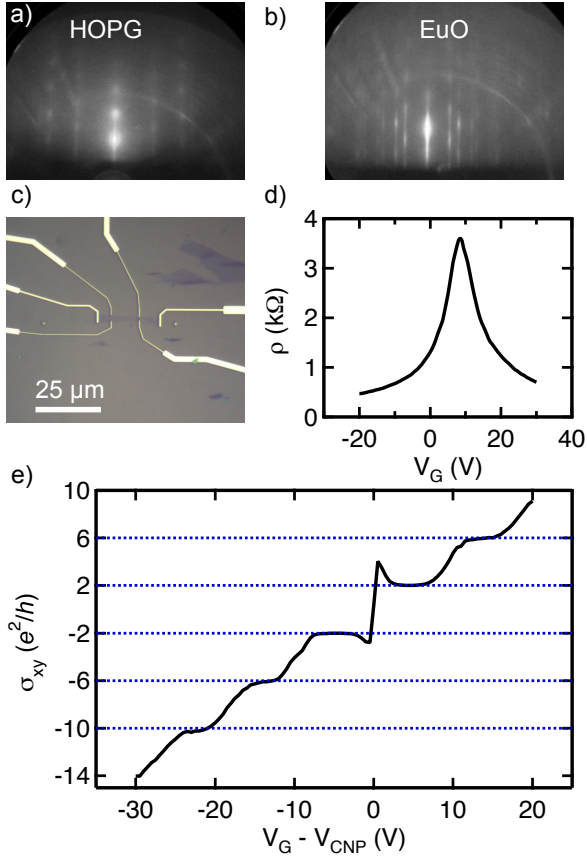}
\caption{\label{fig:1} a) RHEED image of HOPG(0001) substrate b) RHEED image for 5 nm EuO deposited on top of HOPG(0001) in the adsorption-limited growth regime. c) Exfoliated single layer graphene flake with Pd/Ti electrodes in a Hall geometry. d) Graphene resistivity, $\rho$, as a function of the back gate voltage. Maximum $\rho$ corresponds with the charge neutrality point (CNP), sometimes called the Dirac point. Positive (negative) voltages beyond the CNP correspond to electron (hole) type carriers. e) Quantum Hall effect in single layer graphene at T=10 K and B=7 T.
}
\end{figure}

While there have been many theoretical works on global exchange fields in graphene due to a proximity exchange interaction caused by the presence of a ferromagnetic insulator, there are woefully few experimental works on this topic. Recently, the integration of EuO with sp$^2$ carbon materials was demonstrated \cite{Swartz:2012b}. Fig. \ref{fig:1} a) shows the reflection high energy electron diffraction (RHEED) image of highly oriented pyrolitic graphite (HOPG) substrate. HOPG consists of many graphene layers oriented along the (0001) axis and the sheets are coupled by the weak van der Waals force. Since the substrate is well oriented out of the plane, a RHEED pattern can be obtained despite the in-plane disorder associated with HOPG substrate. The surface of HOPG can be considered to be very similar to graphene from a growth point of view. Fresh surfaces of HOPG(001) are obtained through mechanical cleavage followed by a UHV anneal at 600 $^\circ$C. The base pressure of the growth chamber is less than $1\times10^{-10}$ Torr. 

Using a special adsorption-controlled growth regime, thin stoichiometric EuO films can be readily produced by reactive MBE \cite{Ulbricht:2008,Sutarto:2009,Swartz:2010,Swartz:2012a}. An elemental Eu flux is introduced to the HOPG substrate maintained at 550 $^\circ$C. The elevated temperature allows for full distillation of the incident Eu atomic beam. A partial pressure of, P$_{O_2}=1\times10^{-8}$ Torr, converts a portion of the Eu flux to EuO which is then thermodynamically favored to be deposited onto the sample surface. The remaining Eu flux that is not converted, readily re-evaporates. In this way, the stoichiometry of the film can be maintained since the overpressure of Eu atoms avoids the formation of Eu$_2$O$_3$ and the re-evaporation of the excess Eu avoids oxygen vacancies. Fig. \ref{fig:1} b) shows the RHEED pattern for EuO grown on HOPG(0001). Evident in the RHEED pattern are streaks with lattice constants associated with EuO(001) [100] and EuO(001) [110]. Thus, even though symmetry would favor EuO(111)/HOPG(0001), growth by reactive MBE prefers the low energy surface plane of EuO(001). This is not surprising since EuO shares the rock-salt crystal structure with MgO, which is also known for strongly preferring out of plane growth along the (001) axis. 

\begin{figure}[t]
\centering
\includegraphics[scale=1]{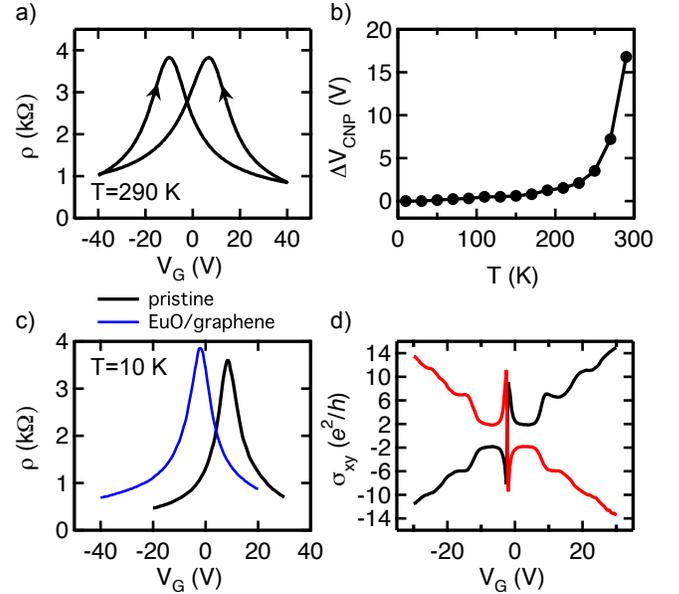}
\caption{\label{fig:2} a) Gate dependent resistivity demonstrating the charging effect at room temperature for EuO deposited on graphene FET device. The back gate is swept from -40 V to +40 V and back again at a rate of 0.5 V/s. b) Temperature dependence of the charging effect plotted as the relative difference in the charge neutrality point between up and down sweeps of the back gate voltage. c) Comparison of the gate dependent resistance at T=10 K for pristine and EuO deposited on top. d) Quantum Hall effect at T=10 K after EuO deposition. Black (red/grey) curve correspond to B=+7T (B=-7 T).}
\end{figure}

Next, we seek to understand the impact of this high temperature growth regime on graphene field effect transistor (FET) devices. Graphene FET devices are fabricated in a Hall geometry as shown in Fig. \ref{fig:1} c). Single layer graphene flakes are mechanically exfoliated onto SiO$_2$(300 nm )/Si substrate and verified by Raman spectroscopy. The heavily p-doped Si substrate acts as the back gate for tuning of the Fermi level. Pd/Ti (60 nm/10 nm) electrodes are fabricated using standard bilayer PMMA/MMA e-beam lithography followed by e-beam deposition and then lift-off. Pd is a significant improvement over Au contacts as the surface diffusion for Pd at elevated temperatures is much less than Au. The Pd/Ti electrodes are specially capped with Al$_2$O$_3$. The device is characterized in a magneto-transport cryostat prior to EuO growth. Gate dependent resistivity measurements at T=10 K are shown in Fig. \ref{fig:1} d). The charge neutrality point is V$_{\text{CNP}}=$ 8 V and the average FET mobility is 5500 cm$^2$/Vs. Pristine graphene FET devices exhibit minimal temperature dependence in the minimum conductivity, mobility, and CNP. Further, these devices exhibit the integer quantum Hall effect as shown in Fig. \ref{fig:1} e). Clear plateaus in the transverse Hall conductance in values of $\pm$2, $\pm$6, and $\pm$10 $e^2/h$ measured at T=10 K and B=7 T is evidence for 2D transport expected for single layer graphene \cite{Novoselov:2005,Zhang:2005}. 

After initial characterization, the device is loaded into the UHV growth chamber and ramped to 550 $^\circ$C at 20 $^\circ$/min. Once the growth temperature is reached, Eu flux is introduced followed quickly by P$_{O_2}$. EuO is allowed to grow for approximately 10 minutes (2 nm) and then the growth is stopped and the sample is cooled back to room temperature at 20 $^\circ$/min. At room temperature, the sample is capped with several nanometers of e-beam MgO. The device is then removed from the UHV chamber, wirebonded \emph{ex-situ}, and then re-loaded into the low-temperature, high-field magnetotransport system. Interestingly, at room temperature, we observe hysteresis in the gate dependent resistivity (see Fig. \ref{fig:2} a)) which is likely due to a charging effect stemming from the dielectric overlayer \cite{Lafkioti:2010,Veligura:2011}. EuO is reported to have a static dielectric constant of 26 \cite{Kasuya:1970a}. As shown in Fig. \ref{fig:2} b) the hysteresis disappears at low temperature. This is quantified by recording the difference in the charge neutrality point, $\Delta \text{V}_{\text{CNP}}$, between up and down gate voltage sweeps. Below approximately 200 K, the difference is less than 1 V. Fig \ref{fig:2} c) compares the gate dependent resistivity at T=10 K and zero field for the pristine device and for 2 nm EuO deposited on top. The CNP has shifted from V$_{\text{CNP}}=8$ V to  V$_{\text{CNP}}=-2$ V and the average mobility has decreased to 3900 cm$^2$/Vs. Further, we still observe clear quantum Hall plateaus in the transverse conductivity, $\sigma_\text{{xy}}$, at the correct values expected for graphene following the equation $\sigma_\text{{xy}} = (m + \frac{1}{2})4e^2/h$, where $m$ is the Landau level filling factor. The fact that the mobility decreases only slightly and the 2D quantum Hall effect is still observed, despite the high temperature deposition of EuO on top an already fabricated device, indicates that the growth does not drastically impact the graphene quality. In particular, the graphene has not functionalized into graphene oxide or some other complex structure.

Anomalous behavior in the Hall effect due to the overlying EuO material is not observed at T=10 K. Observation of such effects which rely on out-of-plane magnetization are likely hindered by the isotropic nature of the ideal Heisenberg exchange which characterizes EuO. Thus, the 2 nm films used here, are highly dominated by shape anisotropy. While the results presented above for thin EuO films on graphene devices is highly promising, integration of the growth process with thicker dielectric films with nanofabrication procedures is necessary. Such advances could potentially lead to the first observation of the quantized anomalous hall effect in single and bilayer graphene \cite{Tse:2011,Qiao:2012}.

\section{Magnetic Moment Formation and Exchange Fields in Graphene Through Dopants and Defects by MBE}
\label{sec:defects}
Induced magnetic phenomena in graphene through adsorbates and defects has been studied extensively theoretically and can be understood in terms of $p_z$-orbital defects \cite{Yazyev:2010}. These defects can be produced through lattice vacancies or through localized sp$^3$ bonding. Experimentally, the majority of the progress has focused on defected graphite and graphene by magnetometry measurements \cite{Yazyev:2010,Esquinazi:2003,Xie:2011,Cervenka:2009,Ramakrishna:2009,Wang:2009,Sepioni:2010,Nair:2012,Nair:2013}. In particular, there have been several reports of ferromagnetic ordering in graphene and graphite \cite{Esquinazi:2003,Xie:2011,Cervenka:2009,Ramakrishna:2009,Wang:2009,Sepioni:2010}. However, theoretically, the ferromagnetic ordering in realistic experiments is not well understood \cite{Yazyev:2010}. Magnetometry measurements have the advantage of directly measuring the total magnetic moment, but are subject to artifacts. The recent work of Nair, et al., \cite{Nair:2012} demonstrated that many HOPG substrates, from which graphene flakes are derived from, are littered with magnetic impurity clusters which could possibly explain the reports of ferromagnetic behavior in magnetometry measurements. The authors went on to conclusively demonstrate spin, S=1/2, paramagnetism induced in graphene laminates through proton irradiation and fluorination. Alternatively, there have been several transport measurements on doped and defected graphene. This has the advantage of locally probing the graphene flake, but does not directly probe the magnetic state in the way that magnetometry does. In 2011, Candini et al., \cite{Candini:2011} observed hysteretic magnetoresistance in narrow graphene ribbons and Hong et al., \cite{Hong:2011} observed negative colossal magnetoresistance and localization in dilute fluorinated graphene. Notably, Chen et al., \cite{Chen:2011} reported the observation of gate tunable Kondo effect in defected graphene through measurements of the temperature dependent resistance which demonstrated a possible logarithmic dependence that could be suggestive of the Kondo effect. This claim was disputed by Jobst and Weber \cite{Jobst:2012}, and Kondo in graphene remains a controversial topic. Very recently, we have demonstrated magnetic moment formation in doped and defected exfoliated graphene flakes by combining the advantages of these two approaches \cite{Mccreary:2012}. More specifically, we seek to probe the spin degree of freedom locally to the graphene in order to minimize the potential for artifacts and minimize the uncertainty in interpretation of the experimental data.

\begin{figure}[t]
\centering
\includegraphics[scale=0.75]{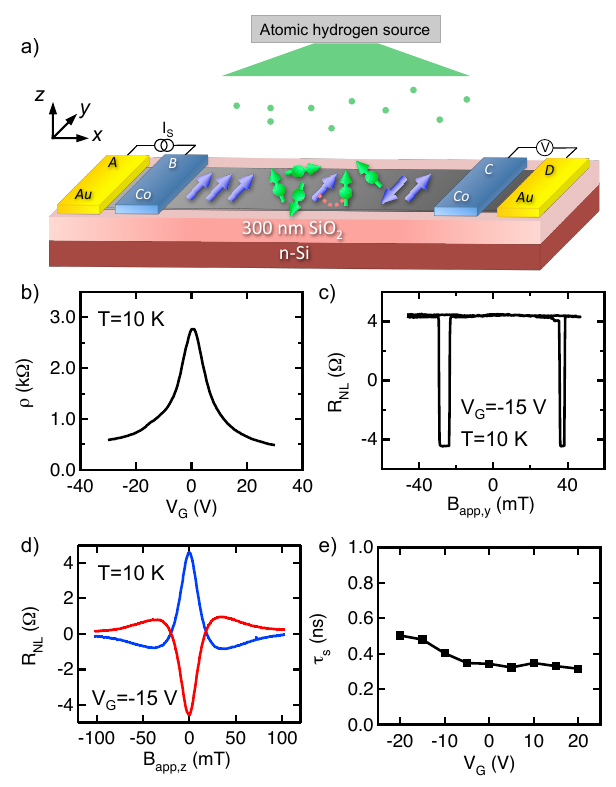}
\caption{\label{fig:3} a) Schematic of the proposed experiment in which atomic hydrogen is carefully introduced \emph{in-situ} to non-local graphene spin valves in an MBE chamber with magnetotransport capability b) Gate dependent resistivity of a pristine non-local graphene spin valve before the introduction of atomic hydrogen. c) Non-local resistance, $R_{NL}$, as a function of in-plane applied field, $B_{app,y}$, along the axis of the magnetic electrodes for pristine graphene. The two state parallel and antiparallel behavior demonstrates spin injection and transport in the graphene channel. Data corresponds with the channel in b) maintained at V$_{\text{G}}=-15$ V. d) Hanle spin precession measurements in pristine graphene at V$_{\text{G}}=-15$ V. d) Gate dependence of the spin lifetime determined from fits to the spin precession data for pristine graphene with tunneling contacts using the Hanle equation. All data is taken at T=12 K.
}
\end{figure}

The proposed experiment is schematically displayed in Fig. \ref{fig:3} a). Non-local spin valves are employed to directly probe the spin degree of freedom locally on the graphene flake. The devices can then be loaded into a UHV deposition chamber (base pressure $1\times10^{-10}$ Torr). The chamber is specially equipped with a sample manipulation stage that is attached to a cryostat. Electrical probes can be connected to the device inside the chamber without exposing the sample to air or needing to vent to manipulate the electrical probes. There is also a home-made Fe core magnet inside the chamber. Thus, magnetotransport measurements on non-local graphene samples can be performed at cryogenic temperatures immediately before and after $p_z$-orbital defects in the form of atomic hydrogen is introduced to the graphene surface. The entire experiment can be performed without needing to warm up or expose the sample to air. Atomic hydrogen is produced from H$_2$ using an Omicron source. High voltage deflectors located in front of the atomic hydrogen source steer charged ions away from the sample. Fabrication and characterization of graphene non-local spin valves with tunneling contacts is reported in detail elsewhere \cite{Han:2010,Han:2011,Han:2012a,Mccreary:2012,Swartz:2013}. Tunneling spin valves, or high resistance contacts, are crucial for this experiment as they alleviate the conductivity mismatch problem and increase the observed spin lifetime in graphene by nearly one order of magnitude \cite{Han:2010,Han:2011,Swartz:2013}.

Fig. \ref{fig:3} b) displays the gate dependent resistivity for a graphene spin valve measured in the four terminal geometry using standard lock-in techniques. The non-local resistance is measured in the configuration displayed in Fig. \ref{fig:3} a). Electrodes B and C are ferromagnetic Co with MgO tunnel barriers, while A and D are spin insensitive Au contacts. Current is passed from electrode B to A, with electrode A held grounded. Spins are injected into the graphene under electrode B. A spatial gradient in the chemical potential causes spins to diffuse to the right along the channel towards electrode C where they can be detected as a voltage relative to electrode D. A non-local resistance, $R_{NL}$ can be defined as the non-local voltage, V$_{NL}$, measured between electrodes C and D, divided by the applied current ($R_{NL}$=$\text{V}_{NL}/\text{I}$). Sweeps of the in-plane magnetic field along the electrode length generates characteristic two state $R_{NL}$ curves as shown in Fig. \ref{fig:3} c) measured at T=12 K and V$_G=-15$ V. 

Instead of applying an in-plane field, we can apply an out-of-plane magnetic field, $B_{app,z}$, in which case the injected spins (oriented in-plane collinear to the magnetization axis of the electrode wires) precess around the applied field. The diffusive transport causes different arrival times at electrode C.  When measured over time scales longer than the spin relaxation and transit time, this appears as dephasing of the spin population with increasing applied field. A typical spin precession measurement taken at T=12 K and V$_\text{G}$=-15 V is shown in Fig. \ref{fig:3} d) for parallel (blue) and antiparallel (red) relative magnetization alignments. For high resistance contacts, the spin precession measurements on pristine graphene are well explained by the Hanle equation\cite{Han:2012a,Mccreary:2012},
\begin{equation}
\label{eq:Hanle1}
R_{NL}\!=\!S_{NL}\!\int \limits_{0}^{\infty}\!\frac{e^{-L^2/4Dt}}{\sqrt{4 \pi D t}}\!\cos\!\left(\!\frac{g \mu_B}{\hbar}B_{app,z}t\!\right)\!e^{-t/\tau_s} dt
\end{equation}	
where $\tau_s$ is of spin lifetime, $D$ is the diffusion coefficient, $S_{NL}$ is the Hanle amplitude, \mbox{$\lambda\!=\!\sqrt{D\,\tau_s}$} is the spin diffusion length and $g$ is the electron $g$-factor. For pristine graphene, $g$ is assumed to be 2. Thus, by fixing the electron $g$-factor, which is well known for pristine graphene\cite{Tombros:2007}, the Hanle data can be uniquely fit using equation \ref{eq:Hanle1} and values for the spin-lifetime and the spin diffusion length can be obtained. In this case, the diffusion coefficient is renamed $D$=$D_S$ to signify that it is determined from the spin measurement. Analysis of the data presented in Fig. \ref{fig:3} d) yields $\tau_s$=479 ps, $D$=$D_S$=0.023 m$^2$/s and $\lambda$=3.3 $\mu$m for the channel length, $L$=5.25 $\mu$m. The spin lifetime for the pristine graphene device is plotted for several gate voltages in Fig. \ref{fig:3} e) and best fits yield spin lifetimes in the range of 300 to 600 ps. Here we distinguish $D_S$ from $D_C$, the diffusion coefficient associated with the charge transport. $D_C$ can be experimentally determined using the Einstein relation, $D_C$=$\sigma/e^2\nu$, where $\sigma$ is the measured channel conductivity, $e$ is the electron charge, and $\nu$ is the density of states (DOS) for single layer graphene. For pristine graphene, $D_C$ and $D_S$ values are expected to agree \cite{Jozsa:2009,Mccreary:2012}

\begin{figure}[t]
\centering
\includegraphics[scale=0.9]{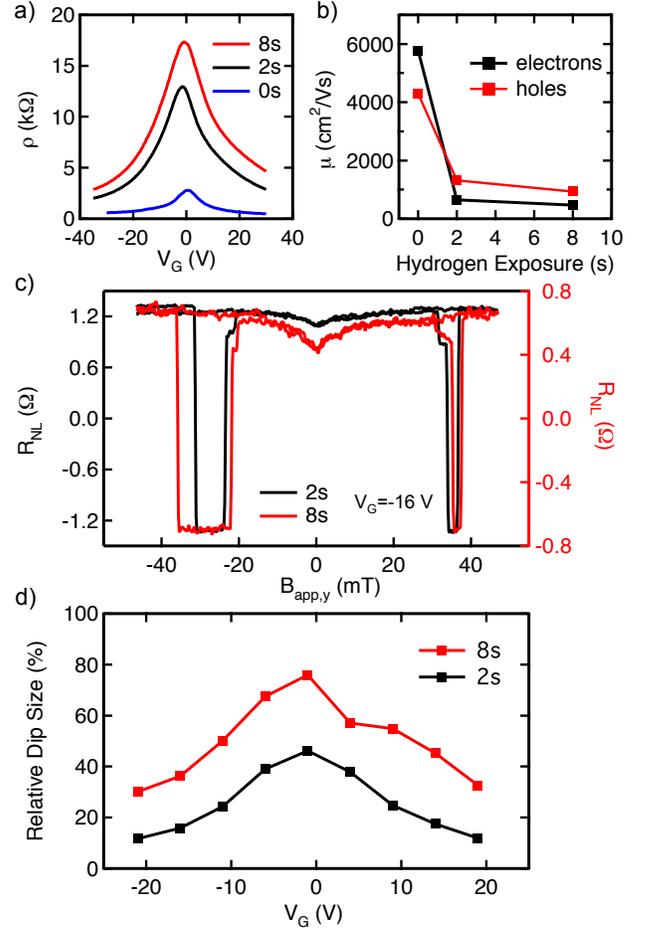}
\caption{\label{fig:4} a) Gate dependent resistivity curves for pristine graphene (blue) and graphene exposed to 2 s (black) and 8 s (red) of atomic hydrogen. The pristine curve (blue) corresponds to the data presented in Fig. \ref{fig:1} b). b) The electron (black) and hole (red) mobilities for the graphene spin valve as a function of hydrogen exposure. c) Non-local resistance, $R_{NL}$ at V$_{\text{G}}=-16$ V for 2 s (left axis, black) and 8 s (right axis, red) of atomic hydrogen exposure. Notably, the dip at zero field increases with increasing hydrogen exposure. d) Gate dependence of the relative dip size compared to the high field $R_{NL}$ values. This is defined quantitatively as $R_{NL}(B_{app,y}\!=\!0) / R_{NL}(B_{app,y}\! \to \! \infty)$. 100\% corresponds with full spin scattering and zero spin signal at zero field.
}
\end{figure}

Next, we examine the effect of atomic hydrogen on charge and spin transport. Atomic hydrogen doping and all measurements occur at T=12 K. Fig. \ref{fig:4} a) displays the gate dependent resistivity for several atomic hydrogen exposures. The blue curve is the pristine device (0 s) and corresponds with the data in Fig. \ref{fig:3} b). The resistivity dramatically increases with 2 s and then again with 8 s hydrogen exposure. The charge neutrality point shifts minimally from V$_{CNP}$=0 V to  V$_{CNP}$=-1 V and the minimum conductivity changes dramatically. This is characteristic of the formation of $p_z$-orbital defects as opposed to charge impurity (CI) scattering \cite{Mccreary:2010,Swartz:2013}. From the reistivity, we can obtain an order of magnitude estimate for the number of $p_z$-orbital defects of 0.1\%.
Accordingly, the mobility is dramatically decreased when the band structure is so affected locally as shown in Fig. \ref{fig:4} b). The decrease in the mobility accompanies a decrease in the diffusion coefficient, which is expected to decrease the overall non-local resistance signal. Thus, as shown in Fig. \ref{fig:4} c) the non-local resistance, $R_{NL}$, decreases with increasing hydrogen exposure for V$_\text{G}$=-16 V. The key feature in the $R_{NL}$ data is the decrease in signal around zero field, called the dip in $R_{NL}$.  With increasing atomic hydrogen exposure, the size of the dip increases relative to the size of the non-local signal at high fields. 

This dip is due to the creation of localized magnetic moments that can interact with the injected spin current via exchange coupling \cite{Mccreary:2012}. A model Hamiltonian including  spin-moment interactions for a single spin interacting with an ensemble of induced moments can be written as,
\begin{eqnarray}
\label{eq:hamiltonian}
H = \eta_M A_{ex} \vec{S}_e \cdot \langle \vec{S}_M \rangle+g \mu_B \vec{S}_e \cdot \vec{B}_{app} \nonumber\\ \\
= g \mu_B \vec{S}_e \cdot \left( \vec{B}_{ex}+ \vec{B}_{app} \right) \nonumber
\end{eqnarray}
where $\eta_M$ is the filling density of magnetic moments, $A_{ex}$ is the exchange coupling between the conduction electron spin ($S_e$) and the spin of the induced moment ($S_M$). The effective field generated by the exchange interaction is given by $\vec{B}_{ex}=\frac{\eta_M A_{ex} \langle \vec{S}_M \rangle }{g \mu_B}$.  In the local frame of the electrons, the exchange field can be described by a time-dependent, randomly fluctuating magnetic field, $\vec{B}_{ex}(t)$, that is the sum of an average mean field term and a randomly fluctuating component,
\begin{equation}
\label{eq:exchangefield}
\vec{B}_{ex}(t)=\overline{\vec{B}}_{ex}+\Delta \vec{B}_{ex}(t).
\end{equation}
These two components of the exchange field impact the observed data in different ways. The exact line-shape of $R_{NL}$ vs. $B_{app,y}$ and the dip are explained by the randomly fluctuating effective exchange field. This fluctuating effective field is characterized by net-zero average value and non-zero rms fluctuations. This is known to generate extra spin relaxation that depends in a characteristic fashion (Lorentzian) upon the applied field ($B_{app,y}$) \cite{Mccreary:2012,Fabian:2007}. On the other hand, the mean field term creates rapid Larmor precession in Hanle measurements, and will be discussed in more detail later in the article. In Fig. \ref{fig:4} d) we show the gate dependence of the relative dip size for 2 s and 8 s hydrogen exposure. This is defined quantitatively as $R_{NL}(B_{app,y}\!=\!0) / R_{NL}(B_{app,y}\! \to \! \infty)$ and 100\% indicates complete scattering (relaxation) of the spin population and zero spin signal at zero field.  The effect on the spin scattering at zero field is symmetric for electrons and holes and the dip size decreases with increasing carrier concentration.

\begin{figure}[t]
\centering
\includegraphics[scale=0.85]{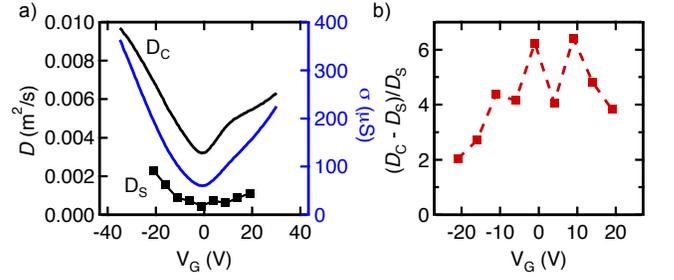}
\caption{\label{fig:5} a) The experimentally determined diffusion coefficient from the Einstein relation ($D_C$) (black curve) and the diffusion coefficient obtained from Hanle fitting with $g$=2 ($D_S$) (black squares) plotted as a function of back gate voltage for 8 s hydrogen exposure. For comparison, the measured conductivity (blue curve) is shown on the right axis. b) Relative difference between $D_S$ and $D_C$ for 8 s hydrogen-doped (dark red squares) graphene.
}
\end{figure}

\begin{figure}[t]
\centering
\includegraphics[scale=0.95]{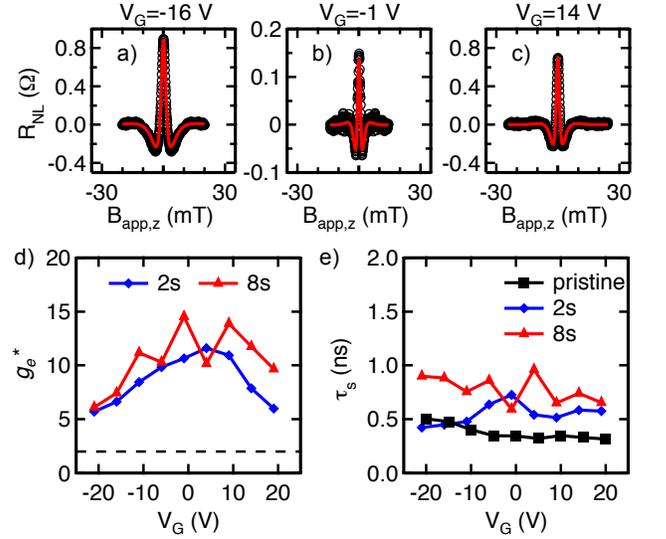}
\caption{\label{fig:6} a) - c) Hanle spin precession curves for graphene spin valves exposed to 8 s atomic hydrogen at several gate voltages. Black open circles represent the measured $R_{NL}$ signal vs. $B_{app,z}$. Red line is the fit to the Hanle equation using fixed $D_C$ (determined from the experimentally measured resistance). The spin lifetime and electron g-factor allowed to be free parameters. d) $g^*$ values determined from Hanle fitting for 2 s (blue diamonds) and 8 s (red triangles) atomic hydrogen exposure. Dashed line indicates $g$=2. e) Spin lifetime obtained from Hanle fitting for pristine graphene (black squares), 2 s hydrogen doping (blue diamonds), 8 s hydrogen doping (red triangles). The pristine data is the same as shown in Fig. \ref{fig:3} e) with $g$=2. The spin lifetimes after hydrogen doping for 2 s and 8 s corresponds with the $g^*$ values shown in d).
}
\end{figure}

We now turn our attention to Hanle spin precession measurements for hydrogen doped graphene. The Hanle curves at V$_\text{G}$=-16 V, -1 V, and 14 V are displayed in Fig. \ref{fig:6} a), b), and c), respectively. For pristine graphene (see Fig. \ref{fig:3} d)), a precession of 180$^\circ$ occurs between 30 mT and 40 mT for all gate voltages. On the other hand, for hydrogen-doped graphene, the Hanle curves are dramatically sharpened with a full 180$^\circ$ rotation occurring within 5 mT at all gates. At first glance, the sharpened Hanle is suggestive that the exchange field is playing a role in the observed spin precession. We now run into a key issue with Hanle analysis. Standard Hanle fitting employs equation \ref{eq:Hanle1} and assumes $g$=2 while allowing the amplitude, lifetime, and diffusion coefficient to be fitting parameters. However, with the possibility that $g\neq2$, then we must replace $g$ with an effective $g$-factor, $g^*$. This is problematic for Hanle fitting since non-unique fits can be obtained with $S_{NL}$, $D$, $\tau_s$, and $g^*$ all free parameters. This can be seen explicitly if we examine the transformation, \mbox{$g^*\rightarrow c\,g^*$}, \mbox{$T_2 \rightarrow T_2/c$}, \mbox{$D_S\rightarrow c\,D_S$}, $S\rightarrow c\,S$, 
where $c$ is a constant. For a given parameter set ($g^*, T_2, D_S, S$), the transformed parameter set ($c\,g^*, T_2/c, c\,D_S, c\,S$) will generate the same Hanle curve. Therefore, if we force a Hanle fit that assumes $g$=2, a transformation that effectively decreases $g^*$ (i.e. $c\!<\!1$) has the effect of artificially enhancing the resulting $\tau_s$ value and artificially decreasing the Hanle diffusion coefficient, $D_S$. This effect is evident in Fig. \ref{fig:5} a) in which the diffusion coefficient for 8 s hydrogen exposure is plotted for both $D_C$ and $D_S$. Here, $D_S$ is obtained from fitting the 8 s hydrogen-doped Hanle data and fixing $g=2$, while $D_C$ is obtained from the measured graphene conductivity using the Einstein relation,
\begin{equation}
\label{eq:Einstein}
D_C = \frac{\sigma}{e^2\nu(E)}.
\end{equation}
Due to charge inhomogeneities in the SiO$_2$/Si substrate, we must use the broadened DOS, $\nu(E)$. For Gaussian broadening \cite{Jozsa:2009}, the DOS is
\begin{equation}
\label{eq:DOS}
\nu(E) = \frac{g_v g_s 2\pi}{h^2v_F^2} \left[ \frac{2b}{\sqrt{2\pi}}e^{-\frac{E^2}{2b^2}}  + E\,\text{erf}\!\left( \frac{E}{b\sqrt{2}} \right)   \right]
\end{equation}
where $g_v$ is the valley degeneracy, $g_s$ is the electron spin degeneracy, $h$ is Planck's constant, $v_F$=1$\times$10$^6$ m/s is the Fermi velocity, and $b$=100 meV is the Gaussian broadening parameter. Accordingly, the calculated $D_C$ shows reasonable agreement with the conductivity (see Fig. \ref{fig:4} a). Unlike the case of pristine graphene \cite{Jozsa:2009,Mccreary:2012}, $D_S$ and $D_C$ are drastically different as summarized in Fig. \ref{fig:5} b). Such a discrepancy is a key indicator that $g^*\neq2$. This an important issue for spintronics as standard Hanle fitting procedures assume the material $g$-factor to be fixed. This is directly relevant to recent experiments concerning $p_z$-orbital defected graphene \cite{Birkner:2013,Wojtaszek:2012}, but has also been discussed for epitaxial graphene on SiC \cite{Maassen:2012,Maassen:2013}. For SiC, the graphene channel may be strongly coupled to localized states in the underlying substrate \cite{Maassen:2013}. In that system, there exists a large discrepancy between $D_C$ and $D_S$, and enhanced effective $g$-factors are observed.

In accordance with the above discussion and following the procedure outlined by \citet{Maassen:2013}, we obtain fits to the Hanle spin precession measurements using equation \ref{eq:Hanle1} with free parameters $S_{NL}$, $\tau_s$, and $g^*$. $D_C$ is fixed to be the experimentally determined value. The fits (red curves) are shown in Fig. \ref{fig:6} a), b), and c) and describe the observed Hanle spin precession data extremely well. In Fig. \ref{fig:6} d) we plot the resulting $g^*$ values as a function of gate voltage for 2 s (blue diamonds) and 8 s (red triangles) hydrogen-doped graphene. The $g^*$ gate dependence is strikingly similar to the relative dip size (see Fig. \ref{fig:4} d)) indicating that the two phenomena are indeed related. This agrees well with the theoretical picture proposed above. With increasing carrier concentration, the effective exchange field, $\vec{B}_{ex}=\frac{\eta_M A_{ex} \langle \vec{S}_M \rangle }{g \mu_B} = \overline{\vec{B}}_{ex}+\Delta \vec{B}_{ex}(t)$, decreases. This impacts both the non-local resistance in plane field measurement and the Hanle spin precession measurements. While the nature of the gate dependence remains an open question, the above equation suggests several factors that could play a role. It is possible that as the Fermi level is tuned away from charge neutrality, that either the $p_z$-orbital defect states become double occupied for positive gates or unoccupied with negative gates. In either case, this would eliminate the local spin moments. This scenario has been supported very recently by magnetometry experiments \cite{Nair:2013}. Alternatively, the exchange coupling may be strongly affected by hybridization or screening, which may further decrease the signal at high carrier concentrations. Here screening may play a significant role by reducing the exchange overlap between a carrier spin traversing a region in the channel and the induced moments at further distances.

The spin lifetime is also obtained through this fitting procedure. Fig. \ref{fig:6} e) displays $\tau_s$ for 2 s (blue diamonds) and 8 s (red triangles) hydrogen exposure. For comparison, the pristine graphene lifetime is displayed (black squares). Notably, there is a slight trend upwards in spin lifetime of roughly a factor of 2 with up to 8 s hydrogen exposure. The average spin lifetimes are 377 $\pm$ 69 ps, 548 $\pm$ 96 ps, and 777 $\pm$ 130 ps, over this back gate range for pristine, 2 s, and 8 s, respectively. This may indicate the importance of the D'yakonov-Perel \cite{DP:1971,Fabian:2007} spin relaxation mechanism in single layer graphene. Unfortunately, this cannot be considered conclusive evidence as the exchange field complicates the analysis. Recent experiments have investigated spin relaxation mechanisms in high mobility graphene \cite{Guimaraes:2012,Neumann:2013} and charge impurity doped graphene \cite{Han:2012b,Swartz:2013}. However, despite the ongoing efforts, this issue remains an open question in the field of graphene spintronics. 

Lastly, we note that this analysis follows a completely different procedure than we employed in \citet{Mccreary:2012} on the same Hanle data. In the previous analysis, the Hanle fitting utilized parameters determined from a unique fitting procedure based on analysis of the dip observed in the in-plane $R_{NL}$ scans as well as involving data taken prior to hydrogen doping. The current analysis is much simpler and does not rely on in-plane $R_{NL}$ scans or data taken prior to hydrogen doping. Importantly, both methods of Hanle analysis yield the same result that $g^*$ is much greater than 2 and has a maximum value near the Dirac point. Because the current Hanle analysis does not involve the in-plane $R_{NL}$ scan, the observation of enhanced $g^*$ provides direct evidence for the formation of magnetic moments that is completely independent of the zero field dip in the in-plane $R_{NL}$ scan.

\section{Conclusion}
Induced spin-based phenomena in graphene is an emerging field with great potential for realizing new physics and device structures. Here we have investigated induced magnetism in graphene and the exchange interaction, which relies on wavefunction overlap. For this, we have employed MBE in order to realize highly controlled interfaces and materials. First, we have examined the integration of EuO onto graphene FET devices in a high temperature adsorbtion-limited regime.  It was found that at low temperatures the device is not dramatically impacted and still behaves nearly as well as the pristine graphene device. This could lead to the future realization of the exchange proximity effect over large regions of a graphene layer. Further, we have examined induced magnetism on a local scale through the systematic introduction of $p_z$-orbital defects in the form of atomic hydrogen. By employing non-local spin valves in a special \emph{in-situ} magnetotransport cryostat and MBE growth chamber, we have been able to systematically investigate the interaction of pure spin currents with the induced moments in a highly controlled environment, without ever needing to warm up the sample or expose to air. Here, the carrier spins interact with the induced moment spin state through the exchange coupling. Effects of this coupling are observed in both in-plane non-local resistance scans and in the Hanle spin precession data. In particular, we have investigated the mean-field component of the exchange field which leads to enhanced effective electron $g$-factors and fast spin precession. The enhanced $g$-factors are gate dependent exhibiting a symmetry around the charge neutrality point with decreasing precession frequency with increasing carrier concentrations. This could possibly be explained by changes in the occupancy of the defect states as the Fermi level is changed or through a screening effect which decreases the coupling between carrier spins and the induced moments. Lastly, here we have proposed a simpler Hanle analysis method compared to a previous treatment of exchange coupling to induced moments in graphene spin valves by \citet{Mccreary:2012}. Notably, this simpler procedure produces the same physical picture and also yields enhanced electron $g$-factors in the range of 5 to 15, well away from the pristine value of 2. These results demonstrate the importance of MBE for investigating spin-based phenomena on the atomic scale.


%
%

%

\begin{acknowledgments}
We acknowledge support from NSF (DMR-1007057,
MRSEC DMR-0820414), ONR (N00014-12-1-0469),
NRI-NSF (NEB-1124601),
NRI-SWAN, ONR (N00014-10-1-0254), and DFG (SFB 689).
\end{acknowledgments}


%

\end{document}